\documentclass{PoS}
\title{
Calculation of $K \to \pi\pi$ decay amplitudes
with improved Wilson fermion
in 2+1 flavor lattice QCD
}
\ShortTitle{
Calculation of $K \to \pi\pi$ decay amplitudes
with improved Wilson fermion action
in 2+1 flavor lattice QCD
}
\author{
\speaker{N.~Ishizuka}${}^{a,b}$\thanks{E-mail : ishizuka@ccs.tsukuba.ac.jp},
K.-I.~Ishikawa ${}^{c  }$,
A.~Ukawa       ${}^{d  }$,
T.~Yoshi\'e    ${}^{a,b}$
\\ \\ \\
\llap{$^a$}
Center for Computational Sciences,
University of Tsukuba, Tsukuba 305-8577, Japan
\\
\llap{$^b$}
Graduate School of Pure and Applied Sciences,
University of Tsukuba, Tsukuba 305-8571, Japan
\\
\llap{$^c$}
Graduate School of Science, Hiroshima University,
Higashi-Hiroshima 739-8526, Japan
\\
\llap{$^d$}
RIKEN Advanced Institute for Computational Science,
Kobe 650-0047, Japan}
\FullConference{
The 32nd International Symposium on Lattice Field Theory  \\
23-28 June, 2014  \\
Columbia University New York, NY
}
\abstract{
We present  results
for the $K\to\pi\pi$ decay amplitudes
for both the $\Delta I=1/2$ and $3/2$ channels.
This calculation is carried out on 480 gauge configurations
in $N_f=2+1$ QCD
generated over 12,000 trajectories with the Iwasaki gauge action and non-perturbatively
$O(a)$-improved Wilson fermion action
at $a=0.091\,{\rm fm}$,
$m_\pi=280\,{\rm MeV}$ and $m_K=580\,{\rm MeV}$
on a $32^3\times 64$ ($La=2.9\,{\rm fm}$) lattice.
For the quark loops in the Penguin and disconnected contributions in the $I=0$ channel,
the combined hopping parameter expansion and truncated solver techniques
work very well for variance reduction.
We obtain, for the first time with a Wilson-type fermion action,  that
${\rm Re}A_0 = 60(36) \times10^{ -8}\,{\rm GeV}$ and
${\rm Im}A_0 =-67(56) \times10^{-12}\,{\rm GeV}$
for a matching scale $q^* =1/a$.
The dependence on the matching scale is weak.
}
\begin{document}
%
%----------------------------------------------------------------------
%
\section{ Introduction }
Calculation of the $K\to\pi\pi$ decay amplitudes for the neutral $K$ meson system
is very important to quantitatively understand the $\Delta I=1/2$ rule
and verify the prediction for the direct $CP$ violation parameter ($\epsilon'/\epsilon$)
in the Standard Model.
However,
a direct lattice calculation of the decay amplitudes
for the $\Delta I=1/2$ process was unsuccessful for a long time,
due in part to a lack of a proper finite-volume formalism,
which has since been laid down in \cite{LL-factor:LL},
and in part to large statistical fluctuations from the disconnected diagrams.
The results of a first direct calculation, on a $16^3\times 32$ lattice,
was reported by RBC-UKQCD collaboration
in Ref.~\cite{A0:RBC-UKQCD_400} at $m_\pi=422\,{\rm MeV}$
with the domain wall fermion action.
They also presented their preliminary results at a smaller quark mass
($m_\pi=330\,{\rm MeV}$) on a $24^3\times 64\times 16$
with the same fermion action
at Lattice 2011~\cite{A0:RBC-UKQCD_300}.

In the present work
we calculate the $K\to\pi\pi$ decay amplitudes
for both the $\Delta I=1/2$ and $3/2$ processes
with the improved Wilson fermion action
on a $32^3\times 64$ lattice with $m_\pi=276 {\rm MeV}$.
Mixings with four-fermion operators
with wrong chirality are absent
for the parity odd process even for the Wilson fermion action
due to the $CPS$ symmetry~\cite{CPS:Donini}.
A mixing to a lower dimension operator does occur,
which gives unphysical contributions to the amplitudes on the lattice.
However, it can be non-perturbatively subtracted
by imposing a renormalization condition~\cite{sub:Dawson}.
After the subtraction one can calculate the physical decay amplitudes
by the renormalization factor
which has the same structure as in the continuum; this is the same situation
as for the domain wall fermion action.
One may expect a gain in the statistical error with the Wilson fermion action,
since it is computationally much less expensive 
than with the domain wall fermion action.

Our calculations are carried out on a subset of
configurations previously generated by PACS-CS Collaboration
with the Iwasaki gauge action and non-perturbatively
$O(a)$-improved Wilson fermion action at $\beta=1.9$
on a $32^3\times 64$ lattice~\cite{conf:PACS-CS}.
The subset corresponds to the hopping parameters
$\kappa_{ud}=0.13770$ for the up and down quark, and
$\kappa_{s }=0.13640$ for the strange quark.
The parameters determined from the spectrum analysis
for this subset are $a=0.091\ {\rm fm}$ and $La=2.91\ {\rm fm}$,
$m_\pi=275.7(1.5)\,{\rm MeV}$ and $m_K=579.7(1.3)\,{\rm MeV}$.
We further generate gauge configurations
at the same lattice parameters to improve the statistics.
The total number of gauge configurations used in the present work is
$480$ which corresponds to $12,000$ trajectories.
We consider the decay of zero momentum $K$ meson  to two zero momentum pions
on these configurations.
The energy difference between the initial $K$ meson
and the final two-pion state takes a non-zero value,
$\Delta E = 21\,{\rm MeV}$ for $I=2$
and $36\,{\rm MeV}$ for $I=0$ on these configurations.
In the present work we assume that
this mismatch of the energy
gives only small effects to the decay amplitudes.
Our preliminary results have been reported
at Lattice 2013~\cite{KPIPI:Our_2013}.
%
%----------------------------------------------------------------------
%
\section{ Calculation }
In the continuum,
the effective Hamiltonian of the $K\to\pi\pi$ decay is given by
a linear combination of 10 four-fermion operators
$Q_i$ for $i=1,2,\cdots 10$~\cite{Review:Buras}.
They can be classified
by the irreducible representation
of the flavor ${\rm SU}(3)_L \times {\rm SU}(3)_R$ symmetry group.
Mixings between operators in different representations are forbidden.
This is also valid for the Wilson fermion action
due to the $CPS$ symmetry as shown in Ref.~\cite{CPS:Donini}
and elaborated in Ref.~\cite{KPIPI:Our_2013}.
However, the mixing to lower dimensional operators has to be considered.
Due to the $CPS$ symmetry and the equation of motion of quark,
there is only one operator with ${\rm dim}<6$, which is
\begin{equation}
   Q_P = ( m_d - m_s ) \cdot P = ( m_d - m_s ) \cdot \bar{s} \gamma_5 d
\ .
\label{eq:Q_P}
\end{equation}
This operator also appears in the continuum,
but  does not give a finite contribution
to the physical decay amplitudes,
since it is a total derivative operator.
This, however, is not valid for the Wilson fermion action
due to the explicit chiral symmetry breaking, and hence
and the operator (\ref{eq:Q_P}) gives a non-zero unphysical contribution
to the amplitudes on the lattice.
This contribution should be subtracted non-perturbatively,
because the mixing coefficient includes
a power divergence of the lattice cutoff as $1/a^2$.
In the present work
we subtract it by imposing the relation~\cite{sub:Dawson},
\begin{equation}
    \langle 0 | \, \overline{Q}_i         \,| K \rangle
 =  \langle 0 | \, Q_i - \alpha_i \cdot P \,| K \rangle = 0
\ ,
\label{eq:sub_Q}
\end{equation}
for each operator $Q_i$.
The subtracted operators $\overline{Q}_i$ are then multiplicatively renormalized
by the renormalization factor having the same form as in the continuum.

\begin{figure}[t]
\begin{center}
\includegraphics[width=100mm]{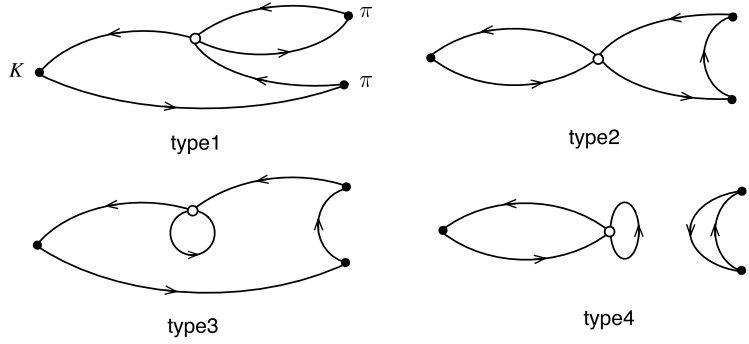}
\end{center}
%
%\vspace{-6mm}
\caption{
Quark contraction of $K\to\pi\pi$ decay.
}
\label{fig:cont}
\end{figure}

We extract the amplitude for each operator
from the time correlation function,
\begin{equation}
  G^{I}_{i}(t)
    = \frac{1}{T} \sum_{\delta=0}^{T-1}\,\,
        \langle 0 | \, W_K(t_K+\delta) \,\, \overline{Q}_i(t+\delta) \,\,
                       W_{\pi\pi}^{I}(t_\pi+\delta) \, | 0 \rangle
\ ,
\label{eq:G_KPIPI}
\end{equation}
where $W_K(t)$ is the wall source for the $K^0$ meson
and $W_{\pi\pi}^{I}(t)$ is that for the isospin $I$ two-pion system.
We impose the periodic boundary condition in all directions.
The summation over $\delta$,
where $T=64$ denotes the temporal size of the lattice,
is taken to improve the statistics.
We set $t_\pi=0$ and $t_K=24$.
We also calculate the amplitudes for $t_K=22$ and $26$
to investigate the ``around-the-world'' effect
which arises from the two-pion operator in the periodic boundary condition
in the time direction.
We confirm that  the effect is small for all channels.
Thus we present results only for $t_K=24$ in the following.
The gauge configurations are fixed to the Coulomb gauge
at the time slice of the wall source $t=t_K+\delta$ and $t_\pi+\delta$
for each $\delta$.
There are four types of quark contractions
for the time correlation function as shown in Fig.~\ref{fig:cont},
where the naming of the contractions follows that
by RBC-UKQCD~\cite{A0:RBC-UKQCD_400}.
%
%---------------------------------------------
%
%
%---------------------------------------------
%
The mixing coefficient of the lower dimensional operator
$\alpha_i$ is evaluated from the ratio,
\begin{equation}
\alpha_i =
\sum_{\delta_1=0}^{T-1} \langle 0 | \, W_K(t_K+\delta_1) \, Q_i(t+\delta_1) \, | 0 \rangle
  \Bigl/ \,
\sum_{\delta_2=0}^{T-1} \langle 0 | \, W_K(t_K+\delta_2) \,   P(t+\delta_2) \, | 0 \rangle
\ ,
\end{equation}
in the large $t_K-t$ region.

For the calculation of the quark loop $Q(x,x)$,  that is
the quark propagator starting from the weak operator
and ending at the same position in the {\it type3 } and {\it type4} contractions,
we use the stochastic method with
the hopping parameter expansion technique (HPE)
and the truncated solver method (TSM)
proposed in Ref.~\cite{HPE_RSM:Bali}.
The detail of application of those method to the $K$ meson decay
has been discussed in Ref.~\cite{KPIPI:Our_2013}.
%
%----------------------------------------------------------------------
%
\section{ Results }
\begin{figure}[p]
\begin{center}
\includegraphics[width=130mm]{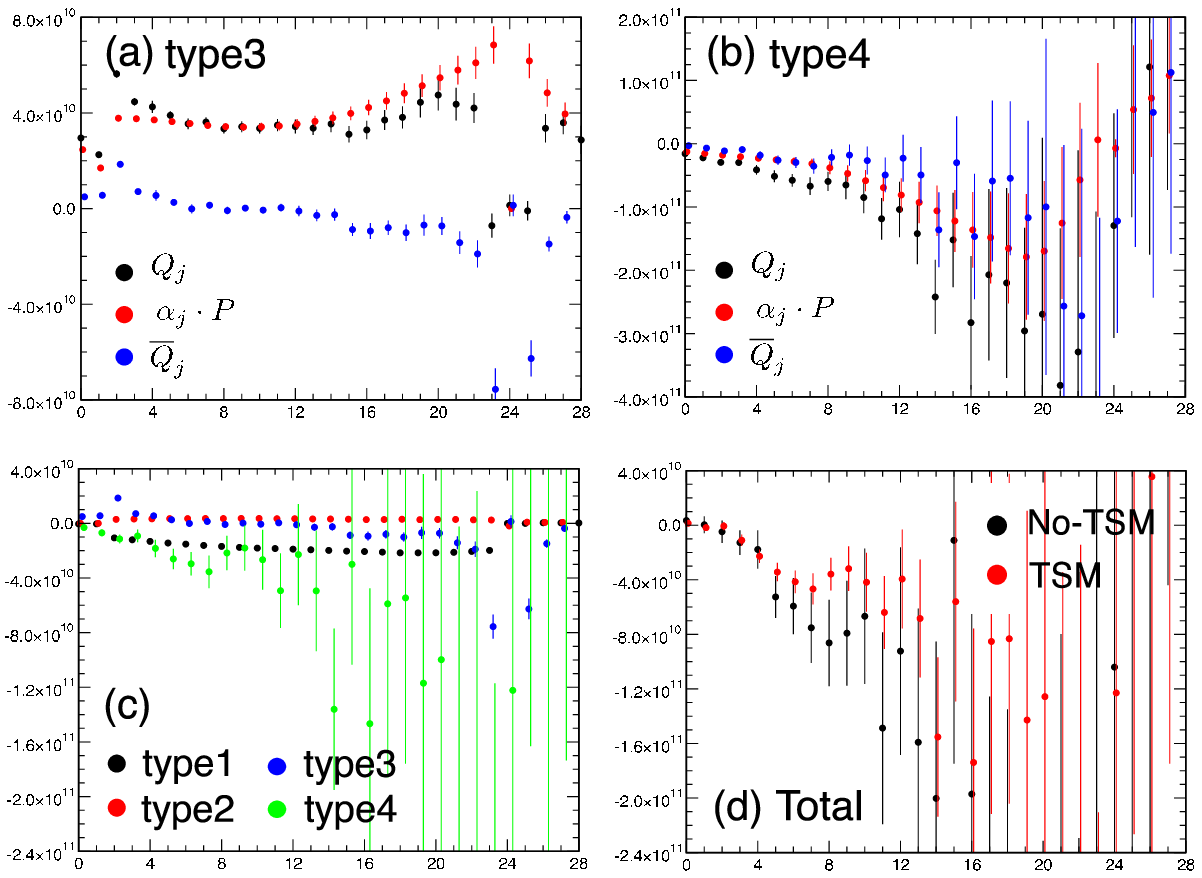}
\end{center}
\vspace{-3mm}
\caption{
Time correlation function
of $Q_2$ for the $\Delta I=1/2$ decay.
(a) {\it type3} contribution for $Q_2$, $\alpha_2\cdot P$ and
     $\overline{Q}_2 = Q_2 - \alpha_2 \cdot P$,
(b)  {\it type4} contribution,
(c) contributions from each type of contractions for $\overline{Q}_2$,
(d) total correlation functions calculated with TSM and without TSM.
}
\label{fig:Q2I0}
%
%--------------------------------------------
%
\begin{center}
\includegraphics[width=130mm]{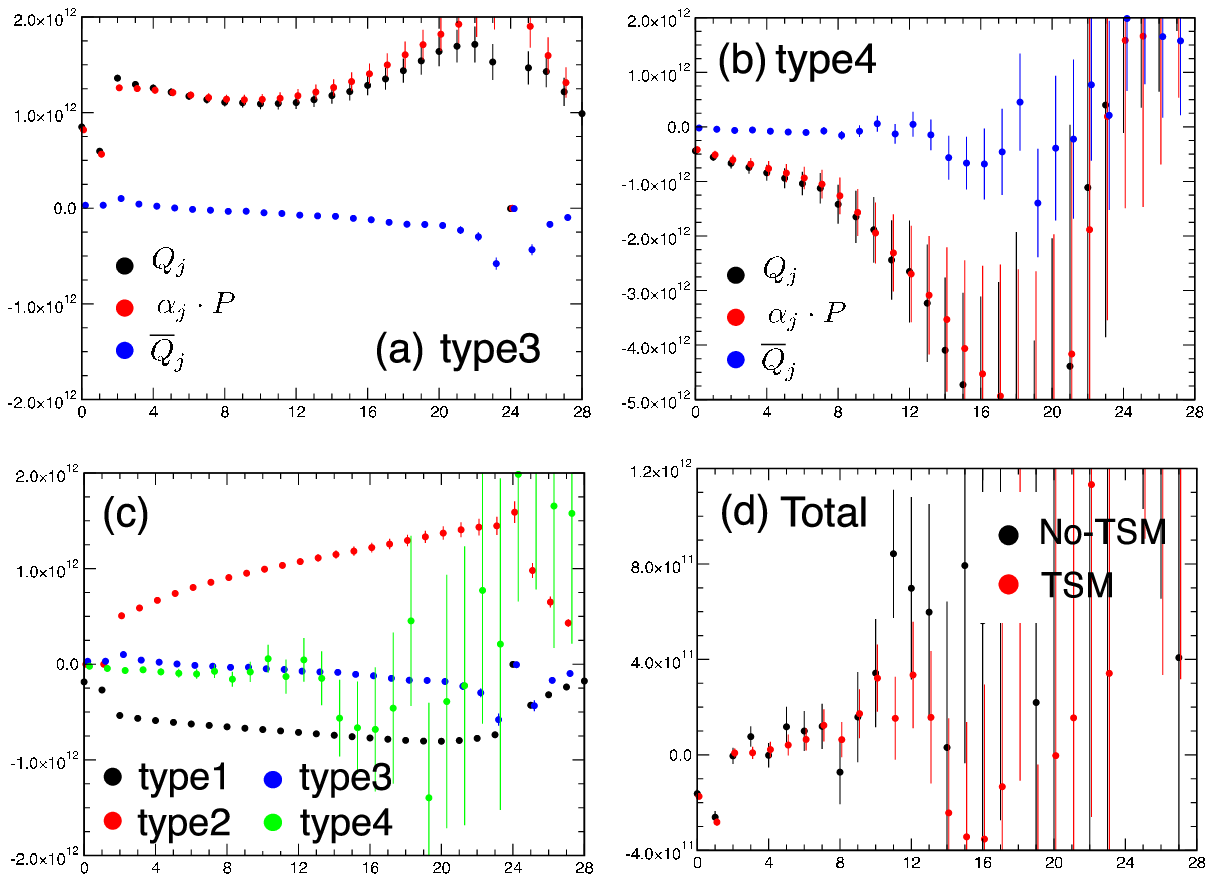}
\end{center}
\vspace{-3mm}
\caption{
Time correlation function
of $Q_6$ for the $\Delta I=1/2$ decay
following the same convention as in Fig.~\protect\ref{fig:Q2I0}.
}
\label{fig:Q6I0}
\end{figure}

The results for the time correlation function
(\ref{eq:G_KPIPI}) of $Q_2$ for the $\Delta I=1/2$ process
are plotted in Fig.~\ref{fig:Q2I0}.
We adopt $K^0 = - \bar{d} \gamma_5 s$ as the neutral $K$ meson operator,
so our correlation function has an extra minus from the usual convention.
We find a large cancellation in $\overline{Q}_2$
between the contributions from the operator $Q_2$ and $\alpha_2\cdot P$
for both the {\it type3} and {\it type4} contractions.
In (c) we find that the contribution from the {\it type4} contraction
is similar in magnitude to that from the {\it type1} contraction.
This appears different from the previous work
by RBC-UKQCD collaboration
with the domain wall fermion action
in Refs.~\cite{A0:RBC-UKQCD_400,A0:RBC-UKQCD_300}.
In (d) we compare the correlation functions
calculated with TSM and without TSM.
We find that TSM significantly improves the statistics.
The numerical cost of TSM is about twice of that without TSM
as was shown in Ref.~\cite{KPIPI:Our_2013}.
Thus TSM is a very efficient method.

The results for $Q_6$ for $\Delta I=1/2$
are plotted in Fig.~\ref{fig:Q6I0}.
Here also  we find a large cancellation in $\overline{Q}_6$ between
the contributions of $Q_6$ and $\alpha_6\cdot P$
for both the {\it type3} and {\it type4} contractions,
as seen for the operator $Q_2$.
In (c) a large cancellation is observed  between
the {\it type1} and  {\it type2} contractions,
which is not the case for $\overline{Q}_2$.
An efficiency of TSM is observed also for $Q_6$ in (d).
%
%---------------------------------------------
%
%---------------------------------------------

We extract the matrix element
$M^I_i = \langle K |\, \overline{Q}_i\, | \pi\pi; I \rangle$
by fitting the time correlation function (\ref{eq:G_KPIPI})
with a fitting function,
\begin{equation}
  G^{I}_{i}(t) = M^{I}_{i}
   / F_{LL} \cdot N_K N_{\pi\pi}^{I}
   \cdot {\rm e}^{ - m_K (t_K-t) - E^{I}_{\pi\pi} (t-t_\pi) }
   \times (-1)
\ ,
\label{eq:amp_M}
\end{equation}
in which  the $K$ meson mass $m_K$ and the energy of the two-pion state $E^{I}_{\pi\pi}$
are fixed at values obtained from the correlation function of the $K$ meson
and the $\pi\pi\to\pi\pi$ process.
The factor $(-1)$ comes from the convention of the $K^0$ operator.
The factors
$N_K = \langle 0 | W_K | K \rangle$ and
$N_{\pi\pi}^{I} = \langle 0 | W_{\pi\pi}^{I} | \pi\pi; I \rangle$
are estimated from the wall to wall propagator of the $K$ meson and the two-pion.
$F_{LL}^{I}$ is the Lellouch-L\"uscher factor~\cite{LL-factor:LL} given by
\begin{equation}
  ( F^{I}_{LL} )^2 = (4\pi)\left( \frac{ E^{I}_{\pi\pi} m_K }{ p^3 } \right)
                           \left( p \frac{ \partial\delta^{I}(p) }{\partial p}
                                + q \frac{ \partial\phi(q) }{\partial q} \right)
 \ ,
\label{eq:F_LL}
\end{equation}
where $\delta^{I}(p)$ is the two-pion scattering phase shift for the iso-spin $I$ channel
at the scattering momentum $p=\sqrt{ E^2/4 - m_\pi^2}$,
and $\phi(q)$ is the analytic function defined in Ref.~\cite{LL-factor:LL}
at $q=p ( 2\pi/L)$.
For the $I=0$ channel the scattering phase shift is not obtained
with a sufficient statistics in the present work.
We leave a precise estimation of the factor to future work,  and
use the value for the non-interacting case,
$(F^{I}_{LL}|_{\rm free} )^2 = (2 m_K L^3 )\cdot ( 2 m_\pi L^3 )^2$,
in the present work.
For the $I=2$ channel we estimate the factor assuming
$\delta^{2}(p) = p (\partial \delta^{2}(p) / \partial p)$
because of the small value of $p$.
We obtain $F^{2}_{LL}/ F^{2}_{LL}|_{\rm free} = 0.9254(62)$.

Our results of the matrix elements for several representative channels are given by
\begin{eqnarray}
&&  a^3 \, M_2^{I=2} = ( 2.256  \pm 0.035 ) \times 10^{-3}  \\
&&  a^3 \, M_7^{I=2} = ( 9.85   \pm 0.11  ) \times 10^{-2} \ , \quad
    a^3 \, M_8^{I=2} = ( 3.242  \pm 0.037 ) \times 10^{-1}  \\
&&  a^3 \, M_2^{I=0} = ( 3.55   \pm 1.32  ) \times 10^{-2} \ , \quad
    a^3 \, M_6^{I=0} = ( - 1.96 \pm 1.06  ) \times 10^{-1} \ ,
\end{eqnarray}
with the lattice constant $a$,
where we adopt $t=[9,12]$ for the fitting range.

The renormalized matrix elements in the continuum
$\overline{M}_{i}^{\,\,I}(\mu)$ are  obtained
from the bare matrix elements on the lattice
$M_j^I$ by multiplying with the renormalization factors,
$\overline{M}_{i}^{\,\,I}(q^*) =\sum_{j} Z_{ij}(q^* a) \, M_j$.
In the present work
we use the renormalization factor
estimated by the tadpole improved perturbation theory in the one loop order
given in Ref.~\cite{Z-fact:Taniguchi}.
We choose two values $q^*=1/a$ and $\pi/a$ as the matching scale
from the lattice to the continuum theory in order to
estimate the systematic error coming from higher orders of  perturbation theory.
The physical decay amplitudes $A_I$ ($I=0,2$) are finally calculated as
\begin{equation}
  A_I = \sum_{ij }\, C_i(\mu)\,  U_{ij}(\mu, q^*) \, \overline{M}_{j}^{\,\,I}(q^*)
      = \sum_{ijk}\, C_i(\mu)\,  U_{ij}(\mu, q^*) \, Z_{jk}(q^* a) \, M_{k}^{I}
\ ,
\end{equation}
where the coefficient functions $C_i(\mu)$
calculated at $\mu=m_c=1.3\,{\rm GeV}$ in Ref.~\cite{Review:Buras} are used.
The function $U(\mu,q^*)$ is the running factor
of the operators $Q_i$ from the scale $q^*$ to $\mu$
for the number of the active fermions $N_F=3$,
which is also given in Ref.~\cite{Review:Buras}.
%
%----------------------------------------------------------
\begin{table}[t]
\begin{center}
\begin{tabular}{l rrrrr}
\hline
%-----------------------------------
& $q^* =1/a$
& $q^* =\pi/a$
& \multicolumn{2}{c}{RBC-UKQCD}
& Exp.
\\
%-----------------------------------
$a\,({\rm fm})$
& \multicolumn{2}{c}{$ 0.091 $}
& $ 0.114 $
& $ 0.114 $
\\
%-----------------------------------
$m_{\pi}\,({\rm MeV})$
& \multicolumn{2}{c}{$ 280 $}
& 330
& 422
& 140
\\
%-----------------------------------
${\rm Re}A_2 \,(\times10^{-8}\,{\rm GeV})$
& $ 2.426(38) $
& $ 2.460(38) $
& $ 2.668(14) $
& $ 4.911(31) $
& $ 1.479(4)  $
\\
%-----------------------------------
${\rm Re}A_0 \,(\times10^{-8}\,{\rm GeV})$
& $ 60(36) $
& $ 56(32) $
& $ 31.1(4.5) $
& $ 38.0(8.2) $
& $ 33.2(2)   $
\\
%-----------------------------------
${\rm Re}A_0 / {\rm Re}A_2$
& $ 25(15) $
& $ 23(13) $
& $ 12.0(1.7) $
& $  7.7(1.7) $
\\
%-----------------------------------
${\rm Im}A_2 \,(\times10^{-12}\,{\rm GeV})$
& $ -1.14(13)   $
& $ -0.7467(83) $
& $ -0.6509(34) $
& $ -0.5502(40) $
\\
%-----------------------------------
${\rm Im}A_0 \,(\times10^{-12}\,{\rm GeV})$
& $ -67(56) $
& $ -52(48) $
& $ -33(15) $
& $ -25(22) $
\\
%-----------------------------------
${\rm Re}(\epsilon'/\epsilon)(\times 10^{-3})$
& $ 0.8(2.5) $
& $ 0.9(2.5) $
& $ 2.0(1.7) $
& $ 2.7(2.6) $
& $ 1.66(23) $
\\
%-----------------------------------
\hline
\end{tabular}
\end{center}
\caption{
Results of the $K\to\pi\pi$ decay amplitudes.
}
\label{table:final_result}
\end{table}
%---------------------------------------------------
%

Our final results for the decay amplitudes are tabulated
in Table.~\ref{table:final_result}.
%, where we show the results
%for the two matching points $q^*=1/a$ and $\pi/a$.
We also list the results
by RBC-UKQCD Collaboration
at $m_\pi=422\,{\rm MeV}$~\cite{A0:RBC-UKQCD_400} and
   $      330\,{\rm MeV}$~\cite{A0:RBC-UKQCD_300},
and the experiment values for comparison.
For ${\rm Re}(\epsilon'/\epsilon)$,
the lattice results for $\epsilon'$ divided by
the experimental value $|\epsilon|=2.228\times 10^{-3}$ are quoted.

We find that the dependence on $q^*$ is negligible for most of the decay amplitudes,
but it is very large for ${\rm Im}A_2$.
Non-perturbative determination of the renormalization factor
is necessary to obtain a  reliable result for this value.

We find a large enhancement of the $\Delta I=1/2$ process over that for $\Delta I=3/2$.
However, our result for $A_0$, particularly for the imaginary part,
still has a large statistical error
so that we do not obtain a non-zero result for ${\rm Re}(\epsilon'/\epsilon)$ over the error.
We observe that the results for $A_0$ by RBC-UKQCD Collaboration
at a similar quark mass $m_\pi=330\,{\rm MeV}$~\cite{A0:RBC-UKQCD_300}
have smaller errors than ours.
This is because they use a different two-pion operator for which the wall sources
for the two pions are separated by $\delta=4$ in the time direction,
and set the fitting range closer to the two-pion source than our case
in extracting the matrix elements from the time correlation function.
Improving statistics by devising some efficient operator
for the two-pion state is an important work reserved for the future.
%
%----------------------------------------------------------------------
%
\section{ Summary }
In the present work
we have reported on our results
of the $K\to\pi\pi$ decay amplitudes
for both the $\Delta I=1/2$ and $3/2$ channels
with the Wilson fermion action.
We have found that the stochastic method with the hopping parameter expansion technique
and the truncated solver method are very efficient for variance reduction,
yielding a first result for the $I=0$ amplitude with the Wilson fermion action.

We have been able to show a large enhancement of the $\Delta I=1/2$ process.
Our result for  $A_0$ and
${\rm Re}(\epsilon'/\epsilon)$ still have large errors, however.
Improving statistics by using some efficient operators
for the two-pion state is necessary to obtain more precise results.

Our calculation is carried out away from the physical quark masses,
and the decay of the K meson to two zero momentum pions is considered.
Calculations at smaller quark masses with physical kinematics,
where the two pions in the final state carry finite momentum, is our next step.
%We must leave these issues to studies in the future.
%
%------------------------------------------------------------------------------
%
\section*{Acknowledgments}
This research used computational resources of the K computer
provided by the RIKEN Advanced Institute for Computational Science
and T2K-TSUKUBA by University of Tsukuba
through the HPCI System Research Project (Project ID:hp120153).
SR16000 at University of Tokyo is also used.
This work is supported in part by Grants-in-Aid
of the Ministry of Education No.~23340054.
%
%---------------------------------------------------------------------------
%

%
%---------------------------------------------------------------------------
%
\end{document}